\documentclass{jpsj-suppl}
\usepackage{txfonts} 

\title{Operational Status and Power Upgrade Prospects 
of the Neutrino Experimental Facility at J-PARC}

\author{Taku \textsc{Ishida} for the T2K Beam Group}

\inst{J-PARC Center, 2-4 Shirane Shirakata, 
Tokai-mura, Naka-gun, Ibaraki 319-1195, Japan 
}

\email{taku.ishida@kek.jp}

\recdate{October 28, 2014}

\abst{
In order to explore CP asymmetry in the lepton sector,
a power upgrade to the neutrino experimental facility at J-PARC 
is a key requirement for both the Tokai to Kamioka (T2K) 
long-baseline neutrino oscillation experiment and 
a future project with Hyper-Kamiokande.
Based on five years of operational experience, the facility 
has achieved stable operation with 230 kW beam power without significant
problems on the beam-line apparatus. 
After successful maintenance works in 2013$-$2014 to replace 
all electromagnetic horns and a production target, 
the facility is now ready to accomodate a 750-kW-rated beam.
Also, the possibility of achieving a few to multi-MW beam 
operation is discussed in detail.
}

\kword{neutrino oscillation, 
neutrino super-beam, 
T2K, 
Hyper-Kamiokande, 
CP violation
}

\begin{document}
\maketitle

\section{Neutrino Experimental Facility at J-PARC}
The neutrino experimental facility at J-PARC was built to produce 
the world's highest intensity neutrino beam 
and to facilitate leading research into neutrino physics. 
Its beam power is rated at 750 kW.
In 2013, the facility was operated at 230 kW, 
which led to the first-ever definitive observation of electron 
neutrino appearance by Tokai-to-Kamioka (T2K)~\cite{t2knueapp}, 
a long-baseline neutrino oscillation experiment between J-PARC 
and Super-Kamiokande (Super-K). 
The observation illustrates that conventional super-beam 
experiments really ``come closer to seeing CP violation''\cite{viewp}: 
T2K has the opportunity to establish 3 $\sigma$ evidence 
of CP violation by accumulating full statistics of data~\cite{futuresens}.
Furthermore, a future project, the long-baseline neutrino oscillation 
experiment from J-PARC to Hyper-Kamiokande (Hyper-K, whose planned 
fiducial volume is $\sim$25 times larger than that of Super-K), 
will allow 5 $\sigma$ observation with an accumulated beam intensity 
corresponding to 7.5 MW$\cdot$year~\cite{hyperK}. 
For these purposes, prompt realization of the designed beam power
is vitally important, and an upgrade to the Mega-Watt beam will 
directly enhance the reach of the future project.

The J-PARC accelerator cascade\cite{JPARCacc} consists of 
a normal-conducting Linac as an injection system, 
a Rapid Cycling Synchrotron (RCS), and a Main Ring synchrotron (MR). 
H$^-$ ion beams, with a peak current of 50 mA and pulse width of 500 $\mu$s, 
are accelerated to 400 MeV by the Linac. 
Conversion into a proton beam is achieved by charge-stripping 
foils at injection into the RCS, which accumulates and accelerates
two beam bunches up to 3 GeV at a repetition rate of 25 Hz. 
Most of the bunches are extracted to the Materials and Life Science 
Facility (MLF) to generate neutron/muon beams. 
The RCS extraction beam power is rated at 1 MW. 
With a prescribed repetition cycle, four successive beam pulses 
are injected from the RCS into the MR to form eight bunches in a cycle 
and accelerated up to 30 GeV. 
In the fast extraction mode, 
the circulating proton beam bunches are extracted 
to the neutrino primary beam-line within a single turn
by a kicker/septum magnet system. 

\begin {figure}[tb]
  \begin{center}
    \includegraphics[width=0.85\textwidth]
     {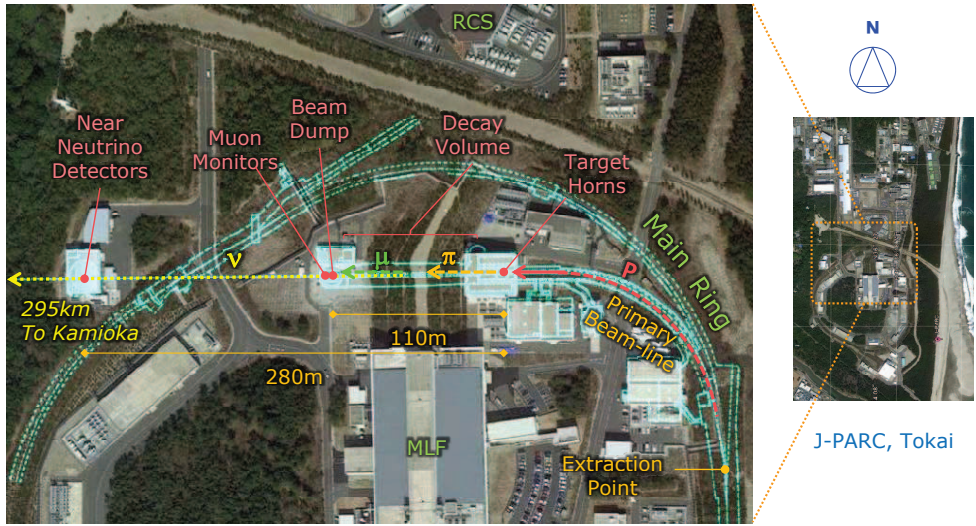}
    \caption{Neutrino experimental facility 
             (neutrino beam-line) at J-PARC.}
    \label{fig:beamline}
  \end{center}
\end {figure}
Fig.~\ref{fig:beamline} shows the layout of 
the neutrino experimental facility\cite{t2k-nim}. 
The primary beam-line guides the extracted proton beam to 
a production target/pion-focusing horn system at
a target station (TS). The focused pions decay into muons and neutrinos 
during their flight through a 110 $m$-long decay volume (DV). 
A beam dump (BD) is located at the end of the DV, 
and muon monitors\cite{mumon} downstream of the BD monitor the muon profile 
on a spill-by-spill basis. 
At 280 $m$ downstream of the target, a neutrino-near-detector 
complex is located, which monitors neutrinos at production.
To generate a narrow-band neutrino beam, 
the beam-line utilizes an off-axis beam configuration\cite{oab} 
for the first time, with the capacity to vary the off-axis angle 
in the range from 2.0$^\circ$ to 2.5$^\circ$. 
The latter value has been used for T2K to date 
and is also assumed to apply to the future project with Hyper-K. 
The centerline of the beam-line extends 295 $km$ to the west, 
passing midway between Super-K and a candidate site for Hyper-K 
($\sim$8 $km$ south of Super-K), so that both sites have identical 
off-axis angles.  

\begin {figure}[tb]
  \begin{center}
    \includegraphics[width=0.7\textwidth]
     {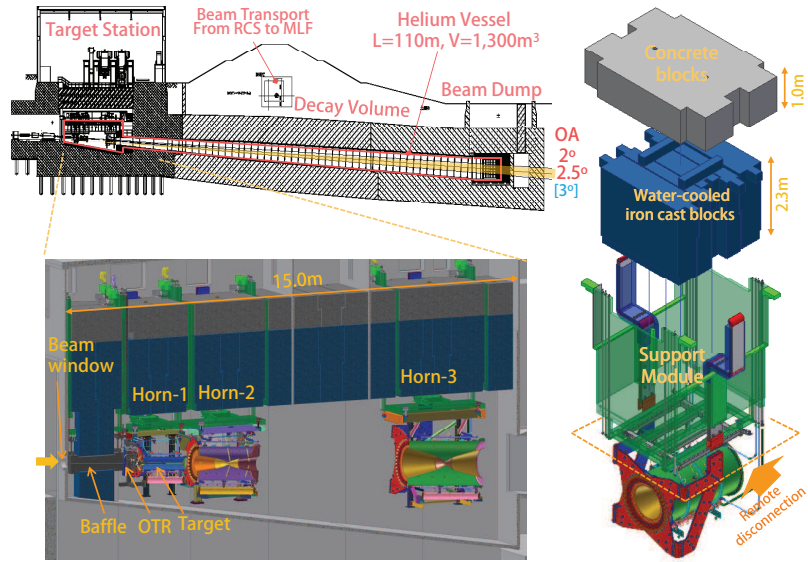}
    \caption{Side view of secondary beam-line, 
             target station, and
             horn support system.}
    \label{fig:secondaryBL}
  \end{center}
\end {figure}

Fig.~\ref{fig:secondaryBL} shows a cross section of 
the secondary beam-line and a close-up of the TS. 
The secondary beam-line comprises the TS, DV, 
and BD. All the apparatus on the beam-axis are enclosed in 
a large vessel of 110 $m$ in length and 1,300 $m^3$ in volume, 
filled with 1 atm of pure Helium gas. 
A beam window separates the helium environment in the 
vessel from the vacuum of the primary beam-line, while
a baffle collimator and three magnetic horns are suspended 
from the walls of the helium vessel by support modules.
A pion production target is inserted into the upstream center of horn-1, 
while an optical transition radiation (OTR) beam profile monitor\cite{otr}
is placed between the baffle and the target. 
Each horn comprises two co-axial cylindrical conductors which can carry 
a 320-kA pulsed current. This generates a peak toroidal magnetic field 
of 2.1 Tesla, which focuses pions with a particular charge.
To remove a heat load from energy deposition by secondary particles 
and from Joule heating, water nozzles assembled into the outer conductor 
spray cooling water onto the inner conductor.
There is also a heat deposition in strip-lines near horns, 
which are surrounded by aluminum ducts and cooled by helium gas 
flowing through them.
The 2.3-$m$-thick cast iron blocks and the 1-$m$-thick concrete 
blocks are also supported by the wall, covering the top of the entire 
target-horn system, for radiation protection. A service pit is located on
top of the concrete blocks, 
where maintenance work by the personnel can be conducted. 
All secondary beam-line components at the TS become 
highly radioactive during operation and replacements require 
handling with a remotely controlled overhead crane.
If, for example, a horn or a target in horn-1 fails, the horn 
(together with its support module) is transferred to a purpose-built 
maintenance area, disconnected from the support module and replaced. 
Failed targets can be replaced within horn-1 using a bespoke target 
installation and exchange mechanism at the maintenance area.

\section{Operational Status of the Neutrino Beam-line}
\label{opst}

\begin {figure}[tb]
  \begin{center}
    \includegraphics[width=0.9\textwidth]
     {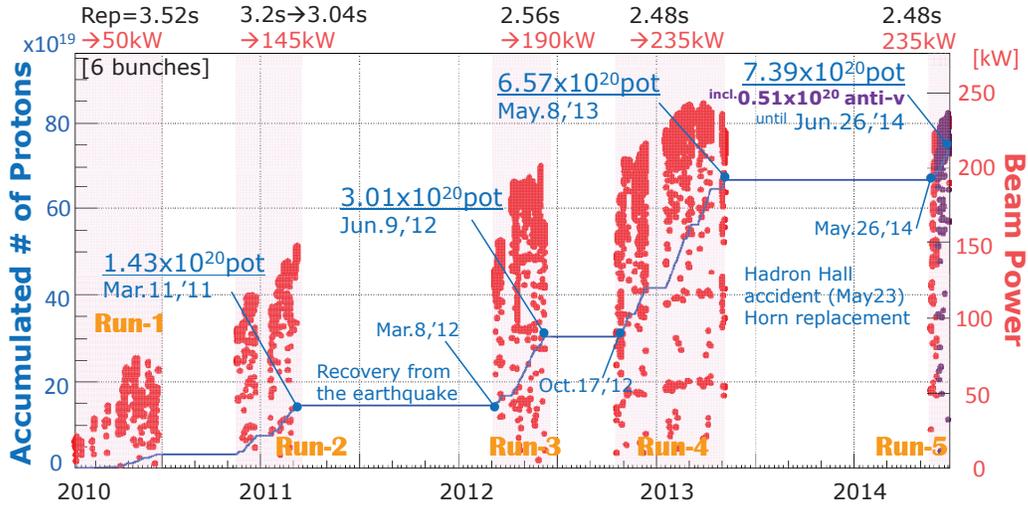}
    \caption{Beam-power history (red dots plotted for each beam pulse) 
     and accumulated numbers of protons-on-target (solid blue line) 
     at the neutrino facility.}
    \label{fig:nupot}
  \end{center}
\end {figure}

Fig.~\ref{fig:nupot} shows the beam-power history 
and total accumulated numbers of protons-on-target (pot)
at the neutrino experimental facility. 
Beam commissioning of the facility began in April 2009 
and the physics data-taking run was initiated in early 2010, 
with 20-kW beam power and a 3.5-s repetition cycle. 
Since then, the beam power has been continuously upgraded 
by increasing the protons-per-pulse (ppp) and shortening the cycle. 
By May 2013, stable operation with $\sim$230-kW beam power 
has been achieved with a repetition cycle of 2.48 s.
The ppp is over 1.2$\times$10$^{14}$ 
(1.5$\times$10$^{13}$$\times$8 bunches), 
which should be noted as the world record for synchrotrons. 
The accumulated pot is 6.57$\times$10$^{20}$ and 
the accumulated number of pulses is 1.2$\times$10$^7$. 
The nominal value of the horn current was set to 250 kA 
and the current stability was within $\pm$5 kA (2 \%) 
for each run period.

Until May 2013, we succeeded in using the first target-horn system 
without serious problems, apart from a 1-year hiatus 
due to the Great East Japan Earthquake.
Meanwhile, during a run beginning in October 2012, 
we detected a substantial problem for the first set of horns: 
Hydrogen gas was being produced inside the conductor by radiolysis 
of the cooling water. After 1 week of 220-kW beam operation, 
the hydrogen concentration was 1.6 \%, which is comparable to the 
explosion limit under atmosphere (4 \%).  
To remove them, a hydrogen-oxygen recombination catalyst, 
composed of alumina pellets with 0.5 \% Pd, was introduced to the system. 
However, since only one port per horn was allocated to fill helium cover gas, 
we were forced to flush/replace the helium using water ports on 
every maintenance day, which was approximately once a week. 
There was also a known limitation for the bus-bar stripline cooling capacity, 
which will be described in Sec.~\ref{horn}. 
In order to be prepared for higher beam power operation, all horns were 
replaced with improved spares during the 2013 maintenance period, 
which was extended to implement safety measures to prevent accidental 
leakage of radioactive material at the hadron hall~\cite{hadacc}. 

A series of replacement works was conducted at the TS 
from September 2013 to April 2014
in a downstream (horn-3) to upstream (horn-1) order. 
The work for each horn was as follows: 
a new horn was docked to a support module under an adjustment stand; 
a test current operation was conducted; 
the old horn was moved from the vessel to the maintenance area;
the new horn was installed in the vessel; and 
the old horn was disassembled from a support module and
disposed of in a casket in the storage area.
This remote exchange scheme worked well, 
and a few hours were required for an old horn to be moved to 
the maintenance area. The residual dose of horn-1 was $\sim$150 mSv/h, 
after 1 year of cooling. While the irradiated horn was being handled 
by the crane, the radiation level at the border of the controlled area 
around the TS was monitored and maintained at a maximum of 4$\mu$Sv/h 
and a few $\mu$Sv in total. These observed values showed good agreement 
with prior simulations by MARS~\cite{MARS}/MCNP~\cite{MCNP} codes.

To prepare for the anti-neutrino mode operation, horn operation with 
inverse polarity ($-$250 kA) was tested, showing good linearity 
and good current balance between bus-bars within a few percent. 
The azimuthal magnetic fields were consistent with those of 
the normal polarity operation. 
After a year of shutdown, a new run was initiated on May 26, 2014. 
The anti-neutrino mode operation began on June 4, 
and the first beam-induced candidate event was observed 
at Super-K on June 8. 
As of June 26, we have accumulated 7.39$\times$10$^{20}$ pot, including 
0.51$\times$10$^{20}$ during anti-neutrino mode operation.

\section{Power Upgrade Prospects}

TABLE~\ref{jparc:MRFXpara} lists the planned parameter sets 
of the MR fast-extraction mode operation, 
with the values achieved as of May 2013: 
Hence, to realize the rated beam power, it is necessary to 
increase both the beam intensity and the repetition rate.
In order to increase the beam intensity, a new accelerating structure, 
ACS (Annular Coupled Structure linac), was installed 
at the downstream end of the Linac during the summer 2013 shutdown, 
and the Linac beam energy was successfully increased from 181 to 400 MeV. 
The front-end component, the ion source and the radio-frequency 
quadrupole (RFQ), were also upgraded during the summer 2014 shutdown, 
to increase the peak current from 30 to 50 mA\cite{linacup}. 
As a result of these upgrades,  
the RCS beam power is expected to reach the rated 1 MW
within FY2015\cite{RCSup}.  
The originally planned MR beam intensity was 3.3$\times$10$^{14}$ ppp,
which is equivalent to 1-MW RCS operation. 
However, a recent tracking simulation shows that the beam loss caused 
by the space charge effect is too large at this intensity,
since the MR physical/collimator apertures are substantially 
smaller than those of the RCS.
To overcome this problem, the accelerator team is following 
a plan\cite{MRmidterm} aiming to ``double the repetition rate'' 
from 1/2.48 to 1/1.28 s. This will be realized in coming years
by replacing the magnet power supplies, 
introducing a new radiofrequency (RF) cavity system with higher gradient, 
and upgrading the injection/extraction devices.
With this scenario, the rated beam power will be achieved with much 
lower beam intensity, {\it i.e.}, 
2.0$\times$10$^{14}$ ppp (610-kW RCS equivalent). 
Furthermore, conceptual studies on how to realize 1- to 2-MW beam powers 
and greater are now under way, for example, 
by increasing the RCS top energy, enlarging the MR aperture, or 
constructing a new ``emittance-damping'' ring between RCS and MR\cite{MRMW}.
\begin{table}[t]
  \caption{Main Ring rated parameters for fast extraction, 
           with numbers achieved as of May 2013.}
  \label{jparc:MRFXpara}
  \begin{center}
    \begin{tabular}{lccc}
      \hline \hline
      Parameter   & Achieved & Original  & Doubled rep-rate \\
      \hline
      Circumference          & \multicolumn{3}{c}{1,567.5\,m } \\
      Beam kinetic energy    & \multicolumn{3}{c}{30\,GeV   } \\
      Beam intensity         & $1.24\times 10^{14}$\,ppp 
      & $3.3\times 10^{14}$\,ppp & $2.0\times 10^{14}$\,ppp \\
      ~                      & $1.57\times 10^{13}$\,ppb 
      & $4.1\times 10^{13}$\,ppb & $2.5\times 10^{13}$\,ppb \\
      $[$ RCS equivalent power $]$  & $[$ 377\,kW $]$ 
      &   $[$ 1\,MW $]$      & $[$ 610\,kW $]$    \\
      Harmonic number        & \multicolumn{3}{c}{9} \\
      Bunch number           & \multicolumn{3}{c}{8~/~spill} \\
      Spill width            & \multicolumn{3}{c}{$\sim$~5\,$\mu$s} \\
      Bunch full width at extraction  & \multicolumn{3}{c}{$\sim$50\,ns} \\
      Maximum RF voltage     & 280\,kV  & 280\,kV & 560\,kV \\
      Repetition period      & 2.48\,sec & 2.1\,sec & 1.28\,sec  \\
       \hline
      Beam power             & 240\,kW & 750\,kW & 750\,kW$\sim$  \\
      \hline
      \hline
    \end{tabular}
  \end{center}
\end{table}

The existing neutrino beam-line 
is designed to accommodate 750-kW beam power, 
while those components that are inaccessible after 
construction/operation, such as the DV, BD, and surrounding concrete shield 
thickness, are designed for future 3- to 4-MW beam power.
It should be noted that, for 750-kW-rated operation, we have assumed 
the original (old) beam parameter set. 
With the lower beam intensity assumed in the doubled rep-rate scenario, 
the thermal shock at both the target and window 
will be reduced significantly. 
As the beam power reaches a few to multi-MW, the beam window, target, and horns 
are expected to become limiting factors, and fundamental modifications 
and/or alternative designs may be required.

%
%
\subsection{Production target}

A production target is inserted within the bore of horn-1, 
as shown in Fig.~\ref{fig:windowtarget}(a). 
It is a 26 $mm$-$\phi$ $\times$ 900-$mm$-long rod of 
isotropic graphite (IG-430U) 
contained within a titanium alloy (Ti-6Al-4V) tube. 
At 750-kW operation, a heat load of approximately 20-kW 
is generated in the target, 
which is removed by 32 g/s helium gas flowing 
over the surface of the rod\cite{cjd-hb2010}. 
The pulsed beam generates an instantaneous temperature increase 
per beam pulse of 200 $^\circ$C and a thermal stress wave 
of magnitude 7.4 MPa, giving a safety factor of $\sim$3.5 
against the tensile strength and allowing for a fatigue factor of 0.9. 
The radiation damage is estimated 
to be $\sim$0.25 displacements per atom (DPA) per year. 
The target has been designed to operate at a maximum temperature 
of $\sim$700 $^\circ$C in order to minimize the radiation damage which 
manifests as dimensional changes and reduction in the thermal conductivity 
of the graphite, as reported from reactor studies\cite{JAERI-rep}. 
On the other hand, the oxidization caused by contamination of the helium gas 
will be increased with this high temperature setting.
Assuming oxygen contamination of 100 ppm
and a graphite temperature of 700 $^\circ$C, 
the safety factor from this effect alone will be reduced 
to $\sim$2 after 5 years of operation\cite{tn-nufact07}. 
When radiation damage effects are considered, 
the maximum lifetime may be significantly less than 5 years. 
By adopting the double rep-rate scenario, 
the instantaneous temperature increase/thermal shock per pulse 
will be reduced to $\sim$60 \%. 
For a beam power of over 750 kW, 
the total proton fluence on the target will be increased, 
thus increasing the heat load and the radiation damage effects. 
The increased heat load can be accommodated by increasing the helium pressure 
and the mass flow rate, although this would have the effect 
of increasing the stress on the target window, which is also affected 
by radiation damage. 
Procedures to re-estimate the maximum acceptable beam power and lifetime 
with respect to the updated beam parameters are now under way.

\subsection{Beam window}

\begin {figure}[tb]
  \begin{center}
    \includegraphics[width=0.8\textwidth]
     {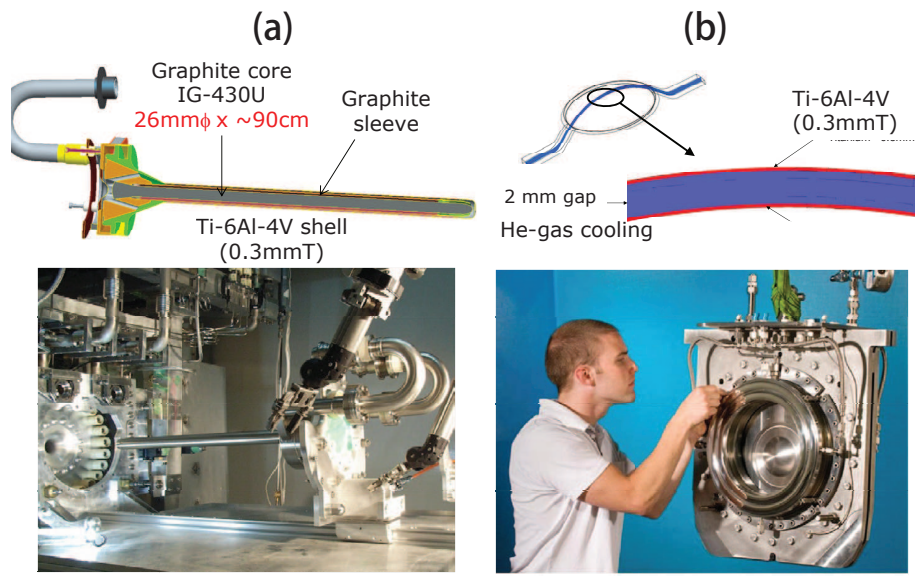}
    \caption{(a) Neutrino production target and 
             (b) beam window at target station.}
    \label{fig:windowtarget}
  \end{center}
\end {figure}

Fig.~\ref{fig:windowtarget}(b) shows a beam window separating 
the helium environment in a TS vessel ($\sim$1 atm) from the vacuum 
of the primary beam-line. In addition to withstanding this differential 
pressure, the window must also survive the thermal stresses 
resulting from interaction with the pulsed proton beam. 
It consists of two 0.3-$mm$-thick concentric partial hemispheres of 
titanium alloy (Ti-6Al-4V, the same material used for the target shell) 
cooled by helium flowing between the two skins (0.08 g/s).
Sealing to the TS vessel and the upstream monitor stack is achieved 
using two inflatable bellows seals.
With the original (old) beam parameters, 
the instantaneous temperature rise/thermal shock are expected to be 
$+$150 $^\circ$C/160 MPa. At thermal equilibrium, the maximum temperature 
and stress are 300 $^\circ$C/200 MPa.
The latter is to be compared with the tensile strength
(1 GPa), which is reduced to 750 MPa at 300 $^\circ$C 
and to 500 MPa after cyclic fatigue. 
The safety factor is, therefore, 2.5.
For the doubled rep-rate scenario with fewer ppp, it will become safer. 
Meanwhile, a reduction of ductility is reported for 
the alloy with 0.10$-$0.24 DPA. 
Based on the number of pot as of 2013, 
the first window at the facility may already have
experienced $\sim$1 DPA. No irradiation data are available for this level, 
which may significantly affect the lifetime estimation.  
In order to increase lifetimes and to permit higher power 
operation for the target and window, a number of studies are planned, 
which include
material irradiation experiments and post-irradiation examination,
studies of alternative target and window materials, and 
the development of conceptual designs/prototyping for segmented 
and packed beds, which would be compatible with the T2K horn 
configuration and associated infrastructure.

\subsection{Electromagnetic horns}
\label{horn}

At 750-kW-rated operation, the inner conductor of horn-1 
will suffer a maximum heat load of 18.5 kJ per pulse, 
assuming a 320-kA current with 2 ms width. 
The aluminium strength, which is moderately reduced by minor 
temperature increases, decreases rapidly by a factor of 4 above 100 $^\circ$C.
For this reason, the tolerable maximum temperature for the conductors 
is set to 80 $^\circ$C. The existing water-spray cooling system has sufficient 
capacity to maintain a temperature of under 60 $^\circ$C, 
and the maximum acceptable beam power is estimated to be $\sim$2 MW.
The new horn system, replaced during the 2013 maintenance, 
have incorporated following improvements:
(1) As mentioned in Sec.\ref{opst}, gaseous hydrogen and 
oxygen are generated from the conductor cooling water through radiolysis. 
For each new horn, an extra tube-line has been added to 
circulate the cover gas through the recombination catalyst system.
This should increase the acceptable beam power to 1$-$2 MW. 
(2) Low inductance and low resistance striplines have been developped 
and introduced, aiming doubled rep-rate operation.
(3) The cooling ducts surrounding the striplines for the first set 
of horns had 20$\sim$30-$mm$ large gaps at the remote-connection 
between the horn and the support module, 
which limited the cooling capacity.
An improved design has been introduced for the new horns, 
and $\sim$1-MW beam power is expected to be achieved with a sufficient helium 
flow rate. For a beam power of more than 1 MW, a fundamental upgrade, 
such as the introduction of a water-cooling scheme, may become necessary. 

So far, two power supplies were used, one for horn-1 
and the other for horn-2 and horn-3 connected in series. 
Meanwhile, in order to realize doubled rep-rate operation, 
we are producing/installing three new power supplies with 
an energy recovery scheme. They are allocated individually 
for each horn to reduce the charging voltage and failure risk.
Another benefit of this upgrade is an increase in the current 
from the nominal 250 kA to the design value of 320 kA.
Owing to these upgrades for the horns and the power supplies, 
the horn system will be ready for 750-kW operation 
with rated 320 kA current within FY2015.

\subsection{Decay volume helium vessel, beam dump, and radiation shielding}
 
Both the DV and BD should dissipate 
$\sim$1/3 of the total beam power, respectively. 
Since neither upgrade nor maintenance of the helium vessel 
are possible after irradiation, they are designed 
to deal with a multi-MW beam (3$-$4 MW).
The DV is composed of 16-$mm$-thick carbon steel plates, 
with 40 water path lines (20 circuits) welded on the inner surface, 
parallel to the beam axis. It is surrounded by 
an approximately 6-$m$-thick concrete wall, 
which can accommodate the radiation due to a 4-MW beam.
A hadron absorber, the core component of the BD, is composed 
of 14 modules of 7 graphite blocks each, clamped to cast aluminium 
water-cooled plates. It is also enclosed in the helium atmosphere. 
Prior to filling with helium gas, the entire vessel is evacuated 
to less than 100 Pa. The impurity of the helium in the vessel is, 
therefore, less than 0.1 \%. 
The oxygen contamination, which is continuously monitored 
during beam operation, is less than 100 ppm. 
The pure helium environment is effective in preventing oxidation 
and production of nitrogen oxides (NOX), which would possibly 
corrode the vessel itself and the internal apparatus.
During the 2013 maintenance, after removing the original horn-3 
from the vessel, a web-camera was inserted to inspect the status of 
the DV and hadron absorber and no damage was detected. 
For 750-kW beam power, the maximum temperature is $\sim$55 $^\circ$C 
for the DV and $\sim$180 $^\circ$C for the hadron absorber. 
The water-cooling systems (pumps, heat exchangers, etc.) 
at the TS and utility buildings only have the capacity for 750-kW operation
and thus must be upgraded to facilitate multi-MW beam operation.

\subsection{Radiation waste treatments and facility upgrades}

When the first T2K physics run began at 20-kW beam power
in early 2010, the air in the service pit was irradiated up to 
$\sim$1 Bq/cc, which escaped between the movable concrete blocks 
and the cables, ducts and piping penetrating the structure. 
As a result, the radiation level in the TS exhaust air reached 
the permitted maximum (0.5 mBq/cc) after only a few hours of 
beam operation. 
Since then, continuous improvements have been made to the overall 
air-tightness, in order to confine the activation underground, 
along with various upgrades to the air ventilation system, 
while the maximum acceptable 
beam power has been reinforced to reach 500 kW$-$1 MW. 
For multi-MW beam operation, further improvements to the air confinement 
and air ventilation system are required.

To maintain water-cooled apparatus (horns, iron blocks, DV, BD, etc.) 
and relevant facilities (water cooling systems, filtration systems, etc.), 
it is necessary to continuously treat the irradiated coolant water.
Although radioactive nuclear ions, such as $^7$Be, 
can be removed using ion-exchange resins, there is no way of 
concentrating the tritiated water (HTO). 
Disposal after dilution, strictly following 
Radiation Hazard Prevention Act, 
is the only way to manage this waste. 
The current water tanks and dilution systems 
in the facility buildings only permit up to $\sim$600-kW beam power,
assuming 100 days of beam operation per year. 
To achieve 750-kW beam operation and to prepare for future multi-MW beams, 
it is necessary to expand the irradiated-water treatment facilities.
Since the floor space of the existing buildings is limited already, 
construction of new utility buildings is highly desirable.        

%
\section{Summary}
\begin{table}[t]
  \caption{Acceptable beam power and achievable parameters 
    for each beamline component after proposed upgrades. 
    Limitations as of May 2013 are also given in 
    parentheses.}
  \label{jparc:BLupgrade}
  \begin{center}
    \begin{tabular}{lcc}
      \hline \hline
      Component   & \multicolumn{2}{c}
      {Acceptable beam power or achievable parameter}  \\
      \hline
      Target      & \multicolumn{2}{c}{3.3$\times$10$^{14}$ ppp } \\
      Beam window & \multicolumn{2}{c}{3.3$\times$10$^{14}$ ppp } \\
      Horn        &   ~  & ~ \\
      \multicolumn{1}{c}{cooling for conductors} & 
      \multicolumn{2}{c}{2 MW} \\
      \multicolumn{1}{c}{stripline cooling}   
      & ( 400 kW $\rightarrow$) & 1$\sim$2 MW \\
      \multicolumn{1}{c}{hydrogen production} 
      & ( 300 kW $\rightarrow$) & 1$\sim$2 MW \\ 
      \multicolumn{1}{c}{power supply} & ( 250 kA $\rightarrow$) & 320 kA  \\
      ~              &  ( 0.4 Hz $\rightarrow$)  &  1 Hz   \\
      Decay volume   &  \multicolumn{2}{c}{4 MW} \\
      Hadron absorber (beam dump)  &  \multicolumn{2}{c}{3 MW} \\
      \multicolumn{1}{c}{water-cooling facilities}  
       & ( 750 kW $\rightarrow$) & $\sim$2 MW \\                  
      Radiation shielding  & ( 750 kW $\rightarrow$) & 4 MW \\
      Radioactive air leakage to the TS ground floor 
      &  ( 500 kW $\rightarrow$) & $\sim$2 MW \\ 
      Radioactive cooling water treatment 
      &  ( 600 kW $\rightarrow$) & $\sim$2 MW \\
      \hline \hline 
    \end{tabular}
  \end{center}
\end{table}
TABLE~\ref{jparc:BLupgrade} gives a summary of the acceptable beam power and/or 
achievable parameters for each beamline component after the proposed upgrades.
Considerable experience has been gained on the path to achieving 230-kW 
beam power operation for T2K, and we are confident that 
the facility is now ready to accommodate a 750-kW rated beam.
Furthermore, we believe the current design/technology of the 
facility is suitable for $\sim$2-MW beam power, in principle,
as the target and beam window were designed for an 1-MW-RCS equivalent beam 
and the current doubled (higher) rep-rate scenario with less ppp causes 
significantly less thermal shock. 
The horns have been replaced with new spares, to which 
many upgrades/improvements have been made, while triple-power-supply 
operation will be realized in the near future, which will make it possible to 
operate horns with a 320-kA-rated current with $\sim$1-Hz repetition.
New utility buildings with larger water tanks for radioactive water drainage 
and higher cooling power facilities are necessary, while
studies on upgrades to the beam-line apparatus and design projects 
for facilities that can accept MW-class beams are now under way.


\begin{thebibliography}{9}
\bibitem{t2knueapp} K. Abe {\em et al.} [T2K collaboration]: Phys. Rev. Lett. 
\textbf{112} (2014) 061802.
\bibitem{viewp} J.A. Formaggio: Physics \textbf{7} (2014) 15.
\bibitem{futuresens} K. Abe {\em et al.} [T2K collaboration]: Submitted to 
Prog. Theor. Exp. Phys., arXiv:1409.7469 [hep-ex].
\bibitem{hyperK} K. Abe {\em et al.} [Hyper-Kamiokande working group]:
P58, presented at J-PARC PAC, May 2014, arXiv:1412.4673 [hep-ex/hep-ph] .
\bibitem{JPARCacc} Y. Yamazaki, K. Hasegawa, M. Ikegami, Y. Irie, 
T. Kato {\em et al.}: J-PARC TDR, KEK-REPORT-2002-13, JAERI-TECH-2003-044, 
J-PARC-03-01.
\bibitem{t2k-nim} K. Abe {\em et al.} [T2K collaboration]: Nucl. Instrum. Meth. 
\textbf{A659} (2011) 106.
\bibitem{mumon} K. Matsuoka {\em et al.}: Nucl. Instrum. Meth. 
\textbf{A624} (2010) 591.
\bibitem{oab} D. Beavis {\em et al.} [E889 Collaboration]: 
BNL preprint BNL-52459 (1995).
\bibitem{otr} S. Bhadra {\em et al.}: Nucl. Instrum. Meth. 
\textbf{A703} (2013) 45.
\bibitem{hadacc} http://j-parc.jp/HDAccident/HDAccident-e.html 
\bibitem{MARS} MARS code system, http://www-ap.fnal.gov/MARS/ .
\bibitem{MCNP} Monte Carlo N-Particle code, https://mcnp.lanl.gov/ .
\bibitem{linacup} K. Hasegawa: This conference.
\bibitem{RCSup} P.K. Saha: This conference.
\bibitem{MRmidterm} T. Koseki: This conference. 
\bibitem{MRMW} S. Igarashi: This conference.
\bibitem{cjd-hb2010} C.J. Densham {\it et al.}: THO2A01, 
Proc. of HB2010, the 46th ICFA Advanced Beam Dynamics Workshop 
on High-Intensity and High-Brightness Hadron Beams, 
Morschach, Switzerland (2010).
\bibitem{JAERI-rep} M. Ishihara {\em et al.}: JAERI-M 91-153 (1991) 
(in Japanese).
\bibitem{tn-nufact07} T. Nakadaira {\it et al.}: CP981, 
Proc. of NuFact07, 9th International Workshop on Neutrino Factories, 
Superbeams and Betabeams, Okayama, Japan (2007). 

\end{thebibliography}
\end{document}